\documentclass[twocolumn,pra,aps,showpacs]{revtex4}
\usepackage{amsmath}
\usepackage{epsfig}

\begin{document}

\title{Irregular Spin Tunnelling for Spin-1 Bose-Einstein Condensates in a Sweeping
Magnetic Field}
\author{Guan-Fang Wang$^{1,2}$ Li-Bin Fu$^1$ and Jie Liu$^{1*}$}
\affiliation{$^1$Institute of Applied Physics and Computational
Mathematics, P.O. Box 8009 (28), 100088 Beijing, China}
\affiliation{$^2$Institute of Physical Science and Technology,
Lanzhou University, 730000 Lanzhou, China}

\begin{abstract}
We investigate the spin tunnelling of spin-1 Bose-Einstein
condensates in a linearly sweeping magnetic field with a mean-field
treatment. We focus on the two typical alkali Bose atoms $^{87}Rb$
and $^{23}Na$ condensates and study their tunnelling dynamics
according to different  sweeping rates of external magnetic fields.
In the adiabatic (i.e., slowly sweeping)  and sudden (i.e., fast
sweeping) limits, no tunnelling is observed. For the case of
moderate sweeping rates, the tunnelling dynamics is found to be very
sensitive on  the sweeping rates with showing a chaotic-like
tunnelling regime.  With magnifying the regime, however, we find
interestedly that the plottings become resolvable under a resolution
of $10^{-4}$ G/s where the tunnelling probability with respect to
the sweeping rate shows a regular periodic-like
 pattern. Moreover, a conserved
quantity standing  for the  magnetization in experiments is found
can dramatically affect the above picture of the spin tunnelling.
Theoretically we have given a reasonable interpretation to the above
findings and hope our studies would bring more attention to spin
tunnelling experimentally.
\end{abstract}
 \maketitle
\section{Introduction}

Bose-Einstein condensation (BEC) has been one of the most active
topics in physics for over a decade, and yet interest in this
field remains impressively high. One of the hallmarks of BEC in
dilute atomic gases is the relatively weak and well-characterized
interatomic interactions. The vast majority of theoretical and
experimental work has involved single component systems, using
magnetic traps confining just one Zeeman sublevel in the ground
state hyperfine manifold, including the BEC-BCS crossover \cite
{regal,zwier}, quantized vortices \cite{matth,madison,raman},
condensates in optical lattices \cite{greiner}, and
low-dimensional quantum gases \cite {olsh,gorlitz}. An impotent
frontier in BEC research is the extension to multicomponent
systems, which provides a unique opportunity for exploring
coupled, interacting quantum fluids. In particular, atomic BECs
with internal quantum structures, some experiments have observed
spin properties of $F=1$ and $F=2$ condensates
\cite{myatt,stenger,miesn,stamp,chang}, using a far-off resonant
optical trap to liberate the internal spin degrees of freedom.
Even $F=3$ bosons are also investigated in a present theoretical
work \cite{hoo}. For atoms in the $F=1$ ground state manifold, the
presence of Zeeman degeneracy and spin-dependent atom-atom
interactions \cite {stenger,stamp1,barrett,ho,law,pu} leads to
interesting condensate spin dynamics, especially spin-1 system
with its relatively simple internal structure. Many literatures
has been devoted to spin mixing and spin domain which had been
observed in experiments.

In this article, we investigate spin tunnelling for spin-1 BEC
with a mean-field description. Unlike all previous studies of the
fixed external magnetic fields both in theory and experiment
(e.g., see Refs \cite{law,pu,wenxian}), we highlight the important
role of an external magnetic field that is now  set to be linearly
varying with time. We focus on the two typical alkali Bose atoms
$^{87}Rb$ and $^{23}Na$ condensates and study their tunnelling
dynamics according to different sweeping rates of external
magnetic fields. We also pay much attention to a conserved
quantity, $m$, standing  for magnetization, and  find that this
quantity can dramatically  affects tunnelling dynamics for both
$^{87}Rb$ and $^{23}Na$ atom system.

Our paper is organized as follows.  Sec.II introduces our model. In
Sec.III we demonstrate our numerical simulations on the irregular
spin tunnelling of alkali Bose atoms $^{87}Rb$ and $^{23}Na$
condensates respectively. In Sec.VI, we present a theoretical
interpretation to the above findings with the  help of both
analytical deductions and  phase space analysis. Sec.V is our
conclusion.

\section{The model}

In an external magnetic field, spin-1 Bose-Einstein condensate (BEC) is
described with the following Hamiltonian \cite{wenxian}
\begin{equation}
H =\int dr[\psi _{i}^{\dagger }(-\frac{\hbar ^{2}}{2M}\nabla
^{2}+v+E_{i})\psi _{i}+ \frac{c_{0}}{2}\psi _{i}^{\dagger }\psi
_{j}^{\dagger }\psi _{j}\psi _{i}+ \frac{c_{2}}{2}\psi
_{k}^{\dagger }\psi _{i}^{\dagger }(F_{\gamma })_{ij}(F_{\gamma
})_{kl}\psi _{j}\psi _{l}], \label{H}
\end{equation}
where repeated indices are summed and $\psi _{i}^{\dagger }(r)$ ($\psi _{i}$%
) is the field operator that creates (annihilates) an atom in the $i$-th
hyperfine state ($\left| F=1,i=+1,0,-1\right\rangle $, hereafter $\left|
i\right\rangle $) at location $r$.\ $M$ is the mass of an atom. Interaction
terms with coefficients $c_{0}$ and $c_{2}$ describe, respectively, elastic
collisions of spin-1 atoms, expressed in terms of the scattering length $%
a_{0}$ ($a_{2}$) for two spin-1 atoms in the combined symmetric channel of
total spin $0$ ($2$), $c_{0}=4\pi \hbar ^{2}\left( a_{0}+2a_{2}\right) /3M$
and $c_{2}=4\pi \hbar ^{2}\left( a_{2}-a_{0}\right) /3M$. $a_{0}$ is not
spin concerned. $a_{2}$ is spin concerned. $F_{\gamma =x,y,z}$ are spin-1
matrices. Assuming the external magnetic field $B$ to be along the
quantization axis ($\widehat{z}$), the Zeeman shift on an atom in state $%
\left| i\right\rangle $ becomes
\begin{equation}
E_{\pm }=-\frac{E_{hf}}{8}\mp g_{I}\mu _{I}B-\frac{E_{hf}}{2}\sqrt{1\pm \xi
+\xi ^{2}}  \label{equation1}
\end{equation}
\begin{equation}
E_{0}=-\frac{E_{hf}}{8}-\frac{E_{hf}}{2}\sqrt{1+\xi ^{2}}  \label{equation2}
\end{equation}
where $E_{hf}$ is the hyperfine splitting and $g_{I}$ is the Lande $g$
factor for an atom with nuclear spin $I$. $\mu _{I}$ is the nuclear magneton
and $\xi =\left( g_{I}\mu _{I}B+g_{J}\mu _{B}B\right) /E_{hf}$ with $g_{J}$
representing Lande $g$ factor for a valence electron with a total angular
momentum $J$. $\mu _{B}$ is the Bohr magneton.

At near-zero temperature and when the total number of condensed atoms (N) is
large, the system can be well described in the mean-field approximation. For
isotropic Bose gas, under the mean-field method and single model
approximation, the operators can be substituted with $c$ numbers $\psi
_{i}=a_{i}\phi (r)$ where $a_{i}$ correspond to the probability amplitudes
of atoms on $i$-th hyperfine state. By setting $a_{i}=\sqrt{s_{i}}e^{i\theta
_{i}}$, the system can be described by the following classical Hamiltionian
system \cite{wenxian},
\begin{equation}
H_{mf} =E_{+}s_{1}+E_{0}s_{0}+E_{-}s_{-1}-c[(1-s_{1}-s_{-1})^{2}
+4s_{1}s_{-1}-4(1-s_{1}-s_{-1})\sqrt{s_{1}s_{-1}}\cos \theta
],\label{H_{mf}}
\end{equation}
where $\theta =\theta _{1}+\theta _{-1}-2\theta _{0}$ and $c=c_{2}\int
dr\left| \phi (r)\right| ^{4}$. Using canonically conjugate transformation, $%
H_{mf}$ can be transfered into the following compact classical Hamitonian
(up to a trivial constant)
\begin{equation}
H_{c}=2cs_{0}[(1-s_{0})+\sqrt{(1-s_{0})^{2}-m^{2}}\cos \theta ]+\delta
(1-s_{0}),  \label{euqation3}
\end{equation}
and equations of motions for canonically conjugate variables $s_{0}$, $%
\theta $ are
\begin{equation}
\dot{s_{0}}=\frac{4c}{\hbar }s_{0}\sqrt{(1-s_{0})^{2}-m^{2}}\sin \theta ,
\label{euqation4}
\end{equation}
\begin{equation}
\dot{\theta}=-\frac{2\delta }{\hbar }+\frac{4c}{\hbar }(1-2s_{0})+\frac{4c}{%
\hbar }\frac{(1-s_{0})(1-2s_{0})-m^{2}}{\sqrt{(1-s_{0})^{2}-m^{2}}}\cos
\theta ,  \label{euqation5}
\end{equation}
where $m=s_{-1}-s_{1}$ is conserved and denoted as magnetization, and $%
\delta =(E_{+}+E_{-}-2E_{0})/2$. Fig.1 shows the relationship between $%
\delta $ and the external magnetic field $B$.
\begin{figure}[tbh]
\begin{center}
\rotatebox{0}{\resizebox *{8.0cm}{8.0cm} {\includegraphics
{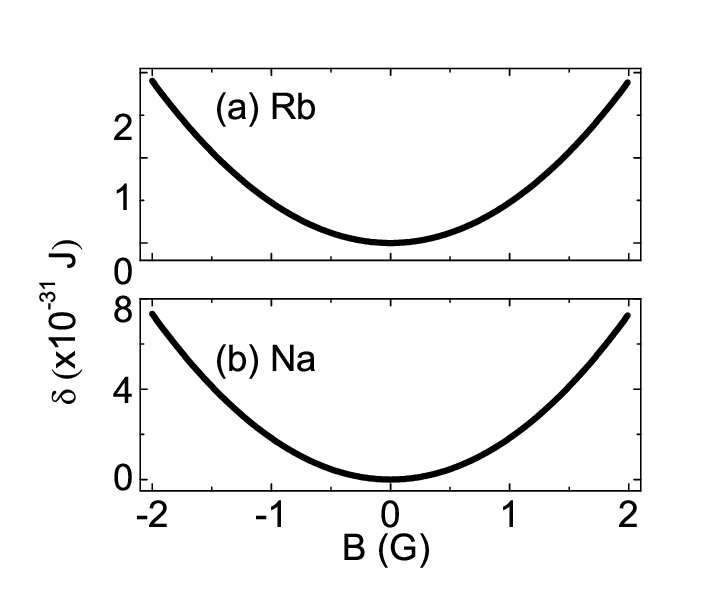}}}
\end{center}
\caption{The transformation of $\protect\delta $, expressed in term of the
Zeeman shift on an atom in state $\left| i\right\rangle $ ($i=+1,0,-1$) in
the external magnetic field $B$. (a) for $^{87}Rb$ atom condensate, (b) for $%
^{23}Na$ atom condensate.}
\label{fig.1}
\end{figure}

\section{Irregular spin tunnelling}

As one of  spin dynamics problems, spin tunnelling is always
interesting to theoretical and laboratorial investigations. In
this section, we study spin tunnelling for the spin-1 BEC systems
in a sweeping magnetic field. In our study, the magnetic field
varies in time linearly , i.e., $\sim \alpha t$, from
$B\rightarrow -\infty $ to $B\rightarrow +\infty $. The sweeping
is far away from Feshbach resonance and ensure that the atom-atom
interaction is almost not variety during the sweeping process.
Numerically, the interval of the magnetic fields is set as
$[-B_0,B_0]$ and the $B_{0}$ is chosen larger enough so that the
coupling between different components are safely ignored at
beginning and ending. In this situation, the spin tunnelling
probability can be well
defined. We want to know the final value of $%
s_{0} $ ( i.e. $s_{0}^{Final}$) at $B\rightarrow +\infty (i.e.,
B_0)$, suppose initially we have  $s_{0}^{Initial}$ at
$B\rightarrow -\infty (i.e., -B_0)$. We exploit Runge-Kutta
$4^{th}-5^{th}$ algorithm to numerically solve the coupled
ordinary
differential equations (\ref{euqation4}) (\ref{euqation5}%
) for the parameters corresponding to Bose atoms $^{87}Rb$ and
$^{23}Na$ respectively.

\subsection{$^{87}Rb$ Atom Condensate}

Because $^{87}Rb$ atom condensate can be readily prepared in
experiments and has been involved in many investigations, we firstly
discus it. For
convenience and not losing generality, we set the initial probability of $%
s_{0}$ as $0.5$ for an example. Fig.2 plots the final value of \ $s_{0}$ at $%
B\rightarrow +\infty $, i.e. $s_{0}^{Final}$ for different sweeping
rates. The above  plotting suggests that our discussions on the spin
tunnelling  can be divided into three parts according to the values
of the sweeping rates  $\alpha $.
\begin{figure}[tbh]
\begin{center}
\rotatebox{0}{\resizebox *{9.0cm}{8.0cm} {\includegraphics
{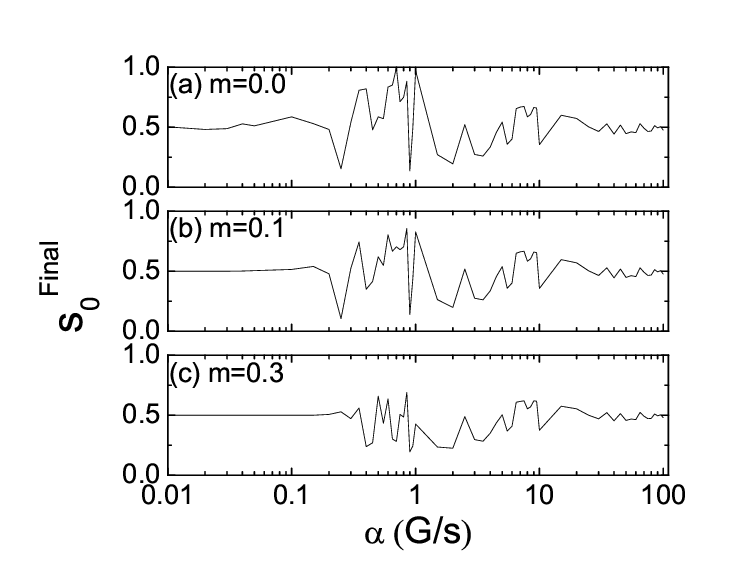}}}
\end{center}
\caption{The tunnelling probability of $^{87}Rb$ for $m=0.0,0.1,0.3$ at $%
c=-3.13\times 10^{-34}J$ in the sweeping magnetic field.}
\label{fig.2}
\end{figure}
\begin{figure}[b]
\begin{center}
\rotatebox{0}{\resizebox *{9.0cm}{9.0cm} {\includegraphics
{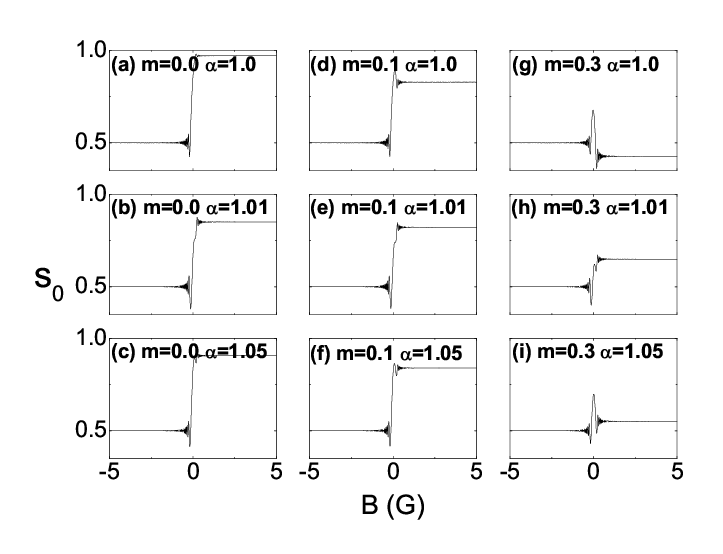}}}
\end{center}
\caption{The change of $s_{0}$ (Its initial is $0.5$ at
$B\rightarrow
-\infty $) with the sweeping magnetic field $B$ for $^{87}Rb$ at $%
m=0.0,0.1,0.3$, $\protect\alpha =1.0,1.01,1.05$ respectively, and $%
c=-3.13\times 10^{-34}J$. } \label{fig.3}
\end{figure}
\begin{figure}[t]
\begin{center}
\rotatebox{0}{\resizebox *{8.5cm}{9.0cm} {\includegraphics
{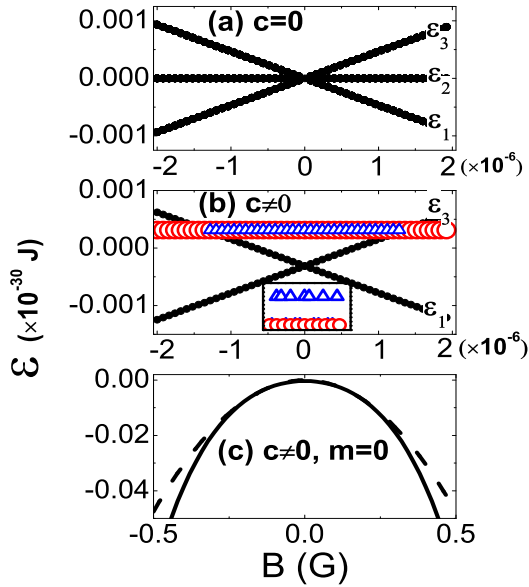}}}
\end{center}
\caption{(Color on line) The eigenvalues of $^{87}Rb$ atom
condensate. (a) is for $c=0$ and different $m$. (b) is for
$c=-3.13\times 10^{-34}J$ and differen $m$, where the triangle is
$\varepsilon _{2a}$ and the circle is $\varepsilon _{2b}$. The
inset shows $\varepsilon _{2a}$ and $\varepsilon _{2b}$ are
non-degenerate. (c) is for $c=-3.13\times 10^{-34}J$ and $m=0$,
where the dash line is $H_{mf1}$ and the solid line is $H_{mf2}$.
To clearly show the spin tunnelling around $B=0$ in fig.3, the
unit of horizons of (a) and (b) is taken to be $10^{-6} G$, while
it is $G$ in (c).} \label{fig.4}
\end{figure}

a) When $\alpha \rightarrow 0$, $s_{0}^{Final}$ almost equal to
initial condition $s_{0}=$ $0.5$, as if the system has not been
changed. The plotting of tunnelling probability vs. sweeping rates
is almost a line, which indicates that no tunnelling occurs after
the external magnetic fields sweep slowly from negative infinity
to positive infinity. In this case, the system is believed to
adiabatically change with the
slowly sweeping magnetic field. When the magnetic field changes from $-\infty $ to $%
+\infty $, $\delta $, the Zeeman shift, changes from an initial
value to zero, then returns  back to its initial value because of
its symmetric
dependence on the field (shown in fig.1). Hence, the classical Hamiltonian (%
\ref{euqation3}) comes back to origin though the field has been
changed dramatically. Numerically solving Eqs.(\ref{euqation4},
\ref{euqation5}) (also see phase space fig.9) we see that for a
fixed magnetic field, the motions are periodic. Hence, in light of
the adiabatic theory\cite{liuliu}, if the sweeping rate of the
external magnetic field is small compared to the frequencies  of
the instantaneous periodic orbits, the
the system undergoes adiabatic evolution. Therefore, $%
s_{0}^{Final}$ $\approx s_{0}.$

b) When $\alpha \rightarrow \infty $, $s_{0}^{Final}$ also tends
to $0.5$ (its inital value). The final value of $s_{0}$ oscillates
around $0.5$ and tends to a line. In this case, the sweeping rate
is so quick, and the system (\ref{euqation3}) restore quickly. If
the time of change the magnetic field is much shorter than the
peroid of the motion of system. It is expected that there is no
time for the system to give some response to the change of field.
\ So no tunnelling phenomenon for very fast sweeping rate can also
be well comprehended.

c) The interesting phenomena emerge when $\alpha $ is moderate. For
this case, we find $s_{0}^{Final}$ changes dramatically with respect
to sweeping rates, which indicates spin tunnelling occurs and the
tunnelling probability is seemingly chaotic\cite{liu}. The spin
tunnelling process can be shown by drawing the evolution of $s_{0}$
with respect to instantaneous magnetic fields $B.$ In fig.3, we plot
the temporal evolution of $s_{0}$ for different sweeping rates
$\alpha =1.0,1.01,1.05$ and for each $\alpha $ we choose  several
magnetization quantities  for comparison. From fig.3, one can read
that the spin tunnelling happens mainly around $B=0$ regardless of
the different quantities of magnetization, and we also see the
tunnelling processes are very sensitive on the sweeping rates.
Moreover we find the conserved magnetization $m$ dramatically
affects the tunnelling processes as well as the final tunnelling
probability. For example, in the first row figures of fig.3, with
increasing the magnetization from 0 to 0.3, we find that the
occupation population of BEC in zero-spin component after a round
sweeping of the external magnetic field changes from being enhanced
to being quenched compared to its initial state. The influence of
the magnetization parameter can be also seen from fig.2, where the
fluctuation on the tunnelling probability is clearly suppressed by
increasing the value of the magnetization.

The crucial  effect of the magnetization parameter on the spin
tunnelling can be roughly understood from eqs.(\ref{euqation4},
\ref{euqation5}). It shows that the variation of the population
$s_0$ is restricted by the conserved magnetization quantity, i.e.,
$|1-s_0|
> |m|$.

To explain why the spin tunnelling happens mainly around zero
value of the magnetic field, i.e., $B=0$, we calculate  the
eigenvalues as well as the eigenstates of the system using the
similar methods developed in our recent work \cite{wang1}. Solving
the eigen equations of the system, we obtain the eigenvalues or
eigenenergies. Fig.4(a) plots them for linear case, i.e., $c=0$.
Fig.4(b) plots them for nonlinear case, i.e., $c\neq 0$. One can
see, at $c=0$, the system has three levels. They are
$\varepsilon_{1}=E_{+}+c, \varepsilon_{2}=E_{0}+c,
\varepsilon_{3}=E_{-}+c$, which cross around $B=0$ and correspond
to $m=-1,0,1$ respectively. When $c\neq 0$, the mid-level
$\varepsilon_{2}$ is split into two levels, i.e.,
$\varepsilon_{2a}=\left( E_{+}+E_{-}\right) /2-c$ (the triangles
in fig.4(b)) and $ \varepsilon_{2b}=E_{0}-c$ (the circles in
fig.4(b)) corresponding to $m=\left(E_{+}-E_{-}\right)/4c$ and
$m=0$, respectively. $\varepsilon_{2a}$ corresponds to the states
of $s_{0}=0$, while $\varepsilon_{2b}$ corresponds to those of
$s_{0}=1$. They are seemingly degenerate, while the inset graph in
fig.4(b) shows they are actually non-degenerate. Furthermore, we
find an interesting phenomenon for this irregular system through
investigating the extreme energies of the classical Hamiltonian.
After adding the nonlinearity to the system, these extremes are
different from the eigenvalues for the same $m$. Taking $m=0$ as
an example, in order to ensure the exact position of the extreme
energies, we use $H_{mf}$ and obtain its two extreme values
$H_{mf1}=E_{0}$, $H_{mf2}=\delta /2+\delta ^{2}/16c+E_{0}+c$. They
are plotted in fig.4(c). Due to these levels are very close around
$B=0$, tunnelling between these levels easily occurs. Moreover,
our calculation reveals that these almost degenerate solutions are
only emerging in the magnetic field of range $\left[-0.16,
0.16\right] G$. It means that the tunnelling should mainly occur
in this regime. The above analysis coincide with the jumping
regime of $s_{0}$ in fig.3. In this way, we explain why the spin
tunnelling mainly happens in a small regime around $B=0$. By the
way, from the above analysis we see that in this system the
eigenstates do not correspond to the extreme values of energy,
e.g., the state of $s_{0}=0.5$ in \cite{wenxian} is not an
eigenstate but a state with an extreme energy. For the same $m$,
tunnelling happens between these extreme energies, while the
eigenstate on the eigenvalue (e.g., $\varepsilon_{2b}$ in
fig.4(b)) is always not change.
\begin{figure}[!tbh]
\begin{center}
\rotatebox{0}{\resizebox *{9.0cm}{9.0cm} {\includegraphics
{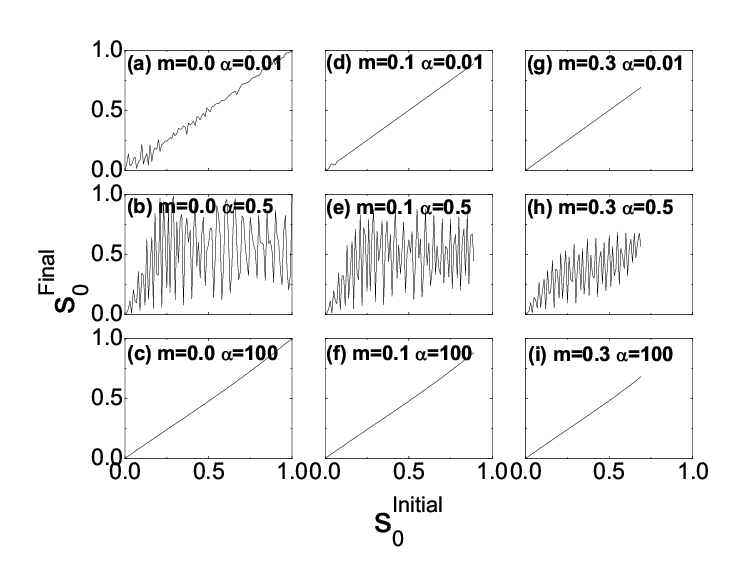}}}
\end{center}
\caption{The relationship between the initial value of $s_{0}$ at $B
\rightarrow -\infty $, $s_{0}^{Initial}$, and its final value at $B
\rightarrow +\infty $, $s_{0}^{Final}$, for $\protect\alpha= 0.01, 0.5, 100$%
, $m=0.0, 0.1, 0.3$ and $c=-3.13\times 10^{-34}J$. } \label{fig.5}
\end{figure}

The above phenomena are general and is independent on the initial
condition, namely $s_{0}.$ To more clearly comprehend this point
of view, we investigate the relationship between the initial value of $s_{0}$ and $%
s_{0}^{Final}$. In fig.5, we take $\alpha =0.01,0.5,100$ as an
example and calculate the relationship between initial $s_{0}$ and
$s_{0}^{Final}$ for several $m$. We see it is a smoothly diagonal
line at $\alpha =0.01,100$, which stands for no tunnelling. At
$\alpha =0.5$, irregular tunnelling occurs. For the same $\alpha
$, the larger $m$, the more smooth the line is, which indicates
that $m$ suppresses the irregular tunnelling.

\subsection{$^{23}Na$ Atom Condensate}

$^{23}Na$ atom condensate has also been prepared in experiments,
and its dynamics are also interesting. Different to $^{87}Rb$, the
interaction between $^{23}Na$ atoms is attractive. In this
subsection, we will discus its spin tunnelling.  In our
discussions, we first  still take the initial probability of
$s_{0}$ as $0.5$ in order to compare with $^{87}Rb$ atom
condensate. Like the above subsection, our discussions are divided
into three parts according to the values of the sweeping rate
$\alpha$. The main results are shown in fig.6. We see the
tendencies of tunnelling probability for $^{23}Na$ atoms are the
same as those of $^{87}Rb$ atoms. Two little differences are found
by comparing the two atom condensates.
\begin{figure}[!tbh]
\begin{center}
\rotatebox{0}{\resizebox *{9.0cm}{9.0cm} {\includegraphics
{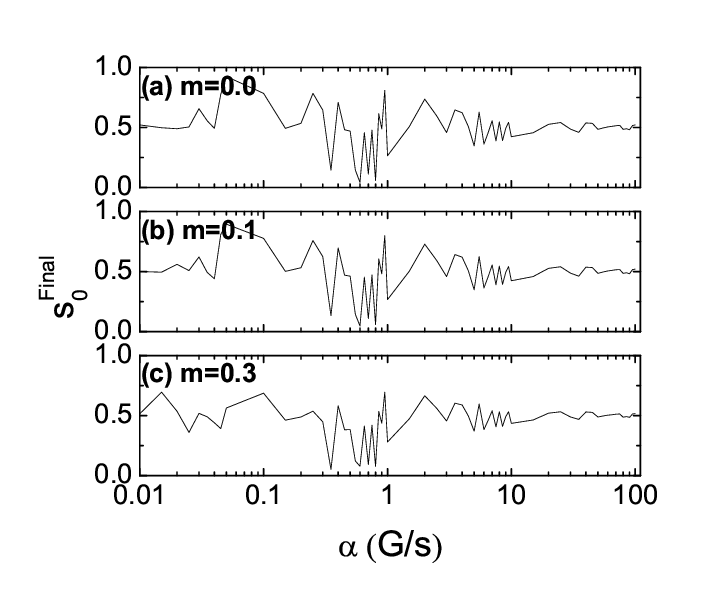}}}
\end{center}
\caption{The tunnelling probability of $^{23}Na$ for $m=0.0, 0.1, 0.3$ at $%
c=3.13\times 10^{-34}J$ in the sweeping magnetic field. }
\label{fig.6}
\end{figure}

The first difference is the adiabatic range. For $^{23}Na$ system,
it is smaller than that of $^{87}Rb$ system. Because the
frequencies of the instantaneous periodic orbits of $^{23}Na$ atom
system is smaller than those of $^{87}Rb$ condensate, the system
needs a less sweeping rate to satisfy the adiabatic condition.
This difference does not affect the fact of no tunnelling
phenomenon at $\alpha \rightarrow 0$. Fig.7 magnifies an
adiabatically chaotic-like part of fig.6(a) and shows the above
phenomenon. When $\alpha \rightarrow \infty$ and $\alpha$ is
moderate, the tunnelling phenomena of $^{23}Na$ system are similar
to the counterparts of $^{87}Rb$ atom system.

The second difference is the effect of the conservation $m$. Like
$^{87}Rb$ atom condensate, the relationship between
$s_{0}^{Final}$ and $s_{0}^{Initial}$ is studied to more clearly
understand the irregular spin tunnelling of $^{23}Na$ atom
condensate. Using the same values of $\alpha$ and $m$ as that in
$^{87}Rb$ atom system, fig.8 shows the relationship. When the
sweeping rate is small, comparing fig.8 with fig.5, we see the
lines in fig.5 are smoother than those in fig.8 for a same
$\alpha$, such as fig.5(d) and fig.8(d). This indicates that the
effect of the conservation $m$ on the spin tunnelling of $^{23}Na$
system is less important. When $\alpha$ is moderate, finite value
of  $m$ suppresses the amplitude of tunnelling probability in
fig.8(b)(e)(h) as well as in fig.5(b)(e)(h). The difference is
that the amplitude of the fluctuation. This can be seen in fig.6
and fig.2. The suppression effect of the finite magnetization $m$
on the irregularity of tunnelling probabilities for $^{23}Na$
system is less significant than that of $^{87}Rb$ system. When
$\alpha$ is larger, i.e., in the sudden limit, fig.8(c)(f)(i) and
fig.5(c)(f)(i) show identical smooth lines without any
fluctuations.
\begin{figure}[!tbh]
\begin{center}
\rotatebox{0}{\resizebox *{9.0cm}{9.0cm} {\includegraphics
{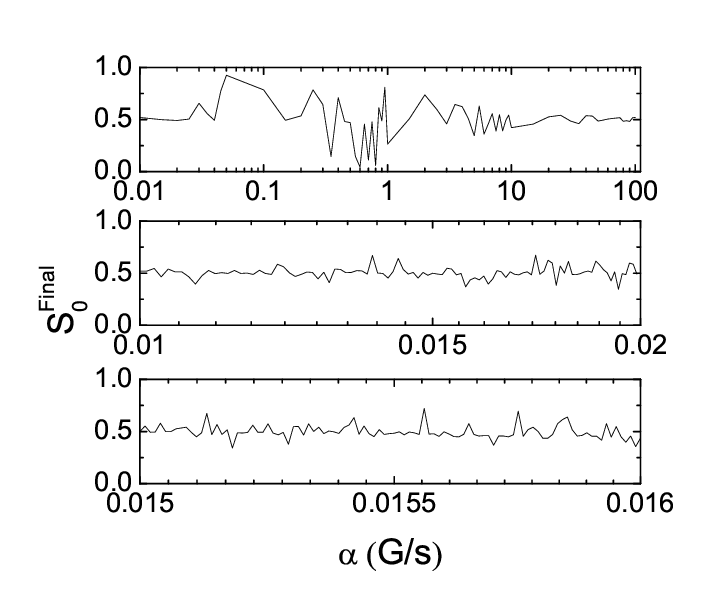}}}
\end{center}
\caption{The magnified part of the tunnelling probability of $^{23}Na$ at $%
m=0.0$, $c=3.13\times 10^{-34}J$ in the adiabatic sweeping
magnetic field. } \label{fig.7}
\end{figure}

\begin{figure}[!tbh]
\begin{center}
\rotatebox{0}{\resizebox *{9.0cm}{9.0cm} {\includegraphics
{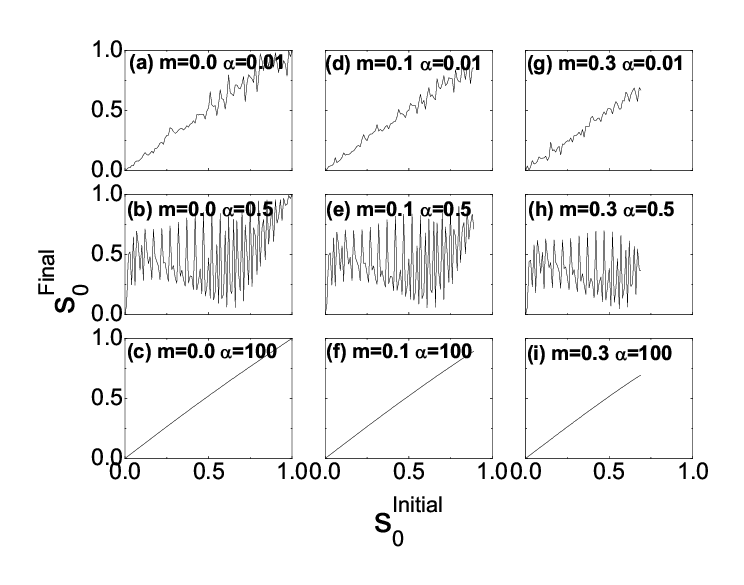}}}
\end{center}
\caption{The relationship between $s_{0}^{Initial}$ and
$s_{0}^{Final}$, for $\protect\alpha= 0.01, 0.5, 100$, $m=0.0,
0.1, 0.3$ and $c=3.13\times 10^{-34}J$. } \label{fig.8}
\end{figure}

\section{Interpretation of the irregular spin tunnelling}

In this section, we achieve insight into the irregular spin
tunnelling effect of spinor BECs with the phase space of the
classical Hamiltonian $H_{c}$\cite{wang2}. As is discussed above,
the tunnelling phenomena of $^{23}Na$ and $^{87}Rb$ atom condensates
have no essential difference. So we take $^{87}Rb$ system as an
example to interpret the irregular spin tunnelling observed in the
above sections.

Fig.9 plots the phase space of Hamilitonian (\ref{euqation3}) for
$^{87}Rb$ for different $\delta /c.$ In these phase space we can
find two different dynamical regions: (I) running phase region where
the relative  phase $\theta $ varies monotonically in time;
(II) oscillation region where $\theta $ oscillates in time around a fixed point. As $%
\delta /c$ varies, the areas of these two regions change
respectively.
\begin{figure}[!tbh]
\begin{center}
\rotatebox{0}{\resizebox *{8.0cm}{7.0cm} {\includegraphics
{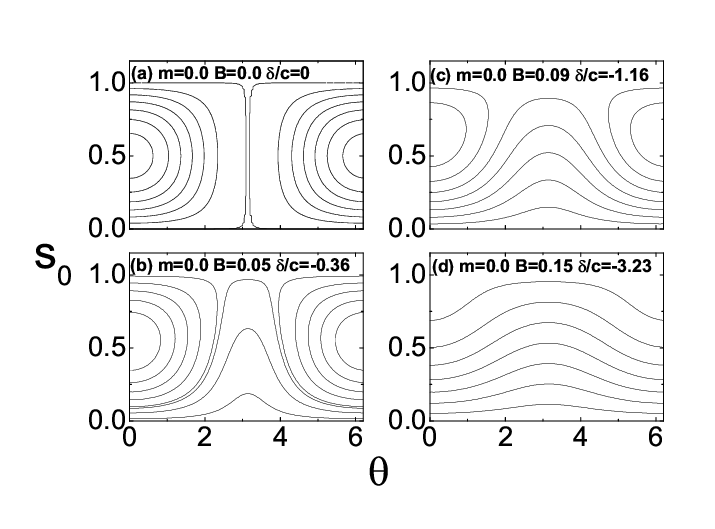}}}
\end{center}
\caption{The phase space of $^{87}Rb$ for different magnetic field $B$ at $%
c=-3.13\times 10^{-34}$J, $m=0.0$. The unit of $B$ is Gauss.}
\label{fig.9}
\end{figure}
\begin{figure}[!tbh]
\begin{center}
\rotatebox{0}{\resizebox *{8.0cm}{7.0cm} {\includegraphics
{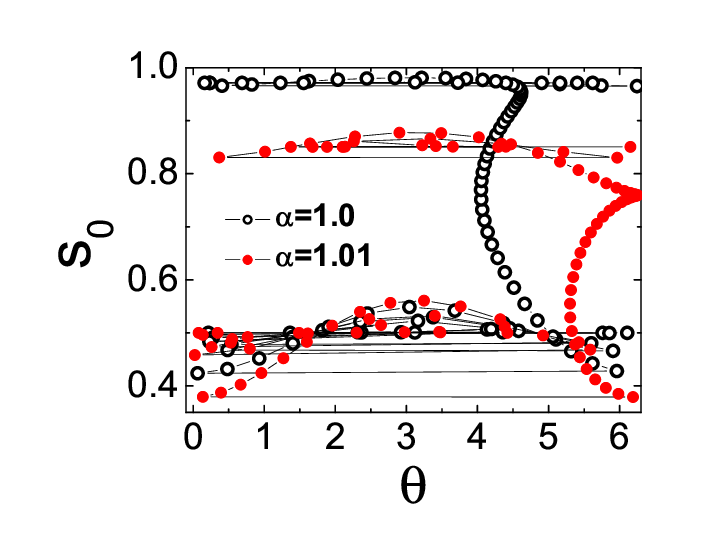}}}
\end{center}
\caption{(Color on line) The phase space of $^{87}Rb$, which is
plotted after an interval of time when the magnetic field moves in
an moderate sweeping rate.} \label{fig.10}
\end{figure}

At first, $\delta /c\sim \infty ,$ all the trajectories are in running phase
region. As $B$ changes with a moderate sweeping rate $\alpha $, we record $%
s_{0},\theta $ after an interval of time and plot them in phase
space (see in fig.10). We see that main contribution to the spin
tunnelling comes from the transition point where the trajectory
passes from the running phase region to the oscillation region or
vice versa. For different $\alpha $, the transition point and
the final equilibrium place are quite different, for example $\alpha =1.0$ and $%
\alpha =1.01$ in fig.10. So the tunnelling probabilities  are
expected to be sensitive on the sweeping rates with showing
irregular patterns observed in Fig.5 and 8.

For higher resolution of $\alpha $, the seemingly chaotic
tunnelling probability is regular. In fig.11 we plot the
magnifying part of fig.2 around $a=1.55$. We find that around a
precision $\alpha $ with high resolution the irregular structure
becomes the regular one having periodic structure. Set its period
is $\alpha_{P}$. When the $\alpha$ is fixed, we find $\alpha_{P}$
has following relation with the initial value of the magnetic
field $B_{0}$: $\alpha_{P} \sim 1/\left|B_{0}\right|^{3}$. Fig.12
shows the relationship between $\alpha_{P}$ and $B_{0}$ around
$\alpha=1.55$ and $\alpha=2.55$. They are
$0.05/\left|B_{0}\right|^{3}$ and $0.15/\left|B_{0}\right|^{3}$
respectively. Furthermore, for a fixed $B_{0}$, the relationship
between $\alpha_{P}$ and $\alpha$ is $\alpha_{P} \sim \alpha^{2}$.
Fig.13 shows $\alpha_{P}=0.0004, 0.0011, 0.0022$ around
$\alpha=1.55, 2.55, 3.55$, respectively. We find
$0.0004:0.0011:0.0022$ is nicely equal to
$1.55^{2}:2.55^{2}:3.55^{2}$. So, the period of regular structure
near a sweeping rate is
\begin{equation}
\alpha _{P} \sim \frac{1}{\left|B_{0}\right|^{3}}\alpha ^{2}
\label{peri}
\end{equation}
\begin{figure}[tbh]
\begin{center}
\rotatebox{0}{\resizebox *{9.0cm}{8.0cm} {\includegraphics
{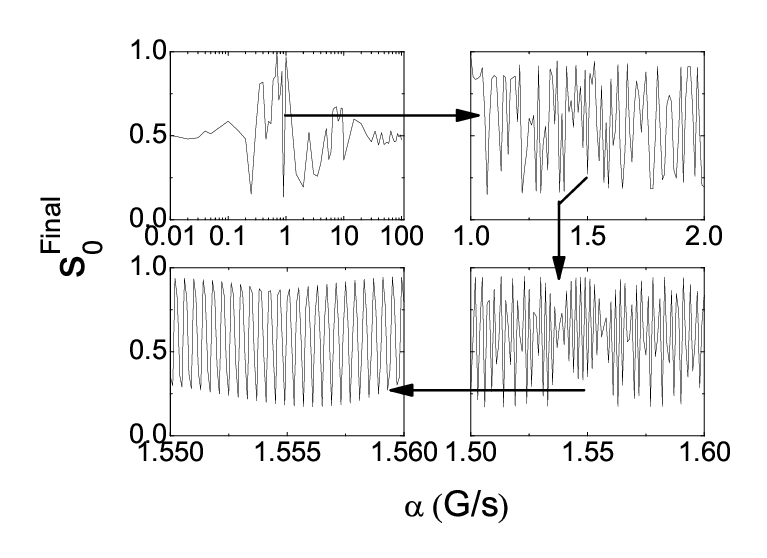}}}
\end{center}
\caption{The magnified part of the tunnelling probability of $^{87}Rb$ at $%
m=0.0$, $c=-3.13\times 10^{-34}J$ in the moderate sweeping
magnetic field. } \label{fig.11}
\end{figure}
\begin{figure}[tbh]
\begin{center}
\rotatebox{0}{\resizebox *{7.0cm}{6.0cm} {\includegraphics
{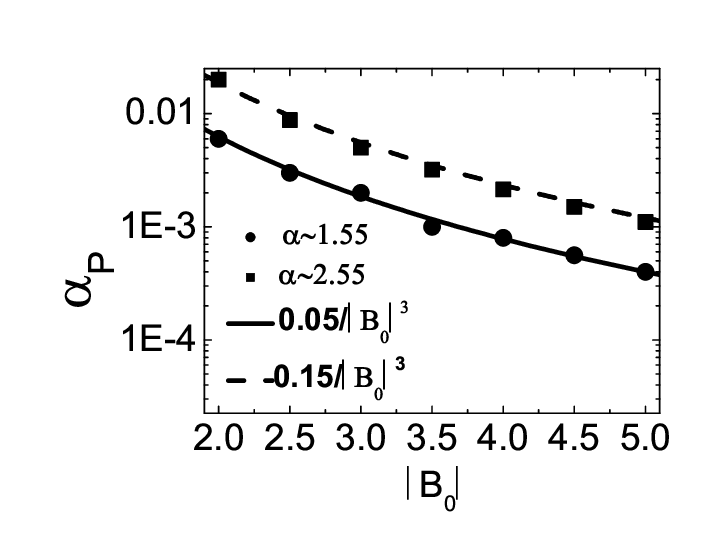}}}
\end{center}
\caption{The relationship between $\alpha_{P}$ and $B_{0}$ around
$\alpha=1.55$ and $\alpha=2.55$. The dots and the squares are
numerical results. The solid line is
$\alpha_{P}=0.05/\left|B_{0}\right|^{3}$. The dash line is
$\alpha_{P}=0.15/\left|B_{0}\right|^{3}$.} \label{fig.12}
\end{figure}
\begin{figure}[tbh]
\begin{center}
\rotatebox{0}{\resizebox *{6.0cm}{9.0cm} {\includegraphics
{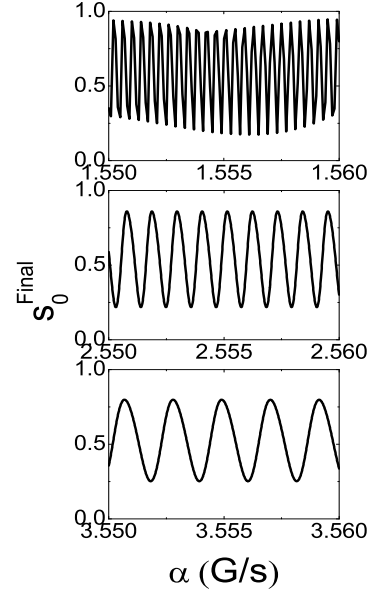}}}
\end{center}
\caption{The magnified part of the tunnelling probability of $^{87}Rb$ at $%
m=0.0$, $c=-3.13\times 10^{-34}J$ around $\alpha=1.55,2.55,3.55$
from up to down, respectively.} \label{fig.13}
\end{figure}
\begin{figure}[tbh]
\begin{center}
\rotatebox{0}{\resizebox *{8.0cm}{9.0cm} {\includegraphics
{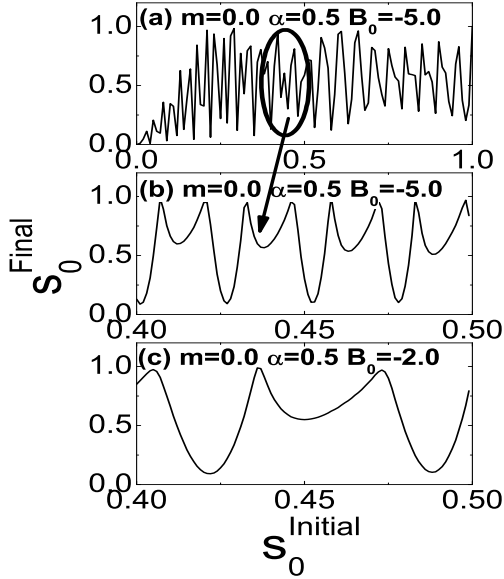}}}
\end{center}
\caption{The magnified part of the relationship between
$s_{0}^{Initial}$ and $s_{0}^{Final}$ around $s_{0}^{Initial}=0.4$
at $m=0.0$, $c=-3.13\times 10^{-34}J$ in a moderate sweeping
magnetic field, i.e., $\alpha=0.5$. (a) is the same as fig.5(b).
(b) and (c) are magnified graph of (a) at $B_{0}=-5.0$ and
$B_{0}=-2.0$ respectively.} \label{fig.14}
\end{figure}
\begin{figure}[tbh]
\begin{center}
\rotatebox{0}{\resizebox *{7.0cm}{6.0cm} {\includegraphics
{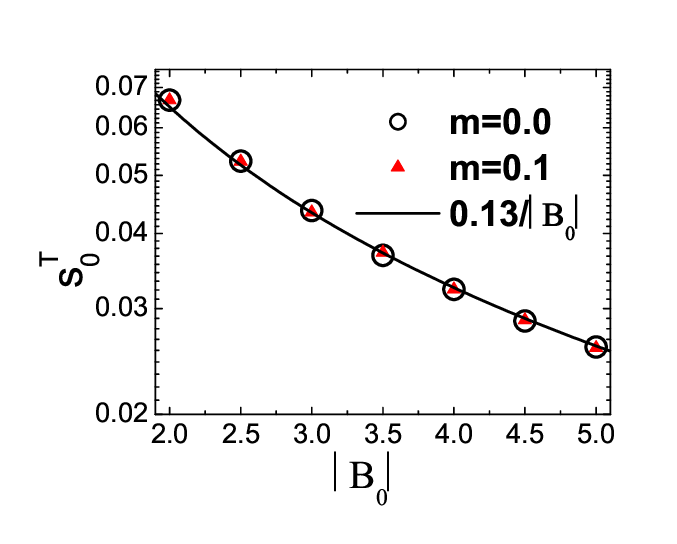}}}
\end{center}
\caption{The relationship between $s_{0}^{T}$ and $B_{0}$ around
$s_{0}^{Initial}=0.4$. The circles and the triangles are numerical
results for $m=0.0, 0.1$, respectively, and $\alpha=0.5$. The
solid line is $s_{0}^{T}=0.13/\left|B_{0}\right|$.} \label{fig.15}
\end{figure}

From this formula, we see that a large $B_{0}$ (used to calculate
tunnelling probability in the above section) leads to a small
$\alpha_{P}$. If the resolution of the sweeping rates is small
compared to the above period, tunnelling probabilities are usually
recorded in different period with random phase. So they will looks
like chaotic. Only when the resolution is high enough compared to
the above period, the regular structure can be observed.

Actually, for a moderate sweeping rate, the chaotic-like
relationship between $s_{0}^{Initial}$ and $s_{0}^{Final}$ in
fig.5 and 8 is also due to the above reason, that is,  when the
resolution of the initial values of $s_{0}^{Initial}$ is
increased, the observed irregular patterns observed are expected
to disappear. Fig.14 shows the magnifying part of fig.5 around
$s_{0}^{Initial}=0.4$. Fig.14(b) and fig.14(c) plot this kind of
regular structure for different initial magnetic field. Setting
the period of this regular structure as $s_{0}^{T}$, we find it
satisfy the following relationship between $s_{0}^{T}$ and
$B_{0}$: $s_{0}^{T} \sim 1/\left|B_{0}\right|$. In fig.15, this
relation is confirmed by our numerical simulations even  for
different magnetization $m$ and a same $\alpha$. So we expect that
the above inversely proportional relation is independent on $m$.

\section{conclusion}

In conclusions, theoretically, we have investigated the tunnelling
dynamics of a spin-1 Bose-Einstein condensate in a linearly
sweeping magnetic field within a framework of mean field
treatment. We focus on the two typical alkali Bose atoms $^{87}Rb$
and $^{23}Na$ condensates and study their tunnelling dynamics
according to different sweeping rates of external magnetic fields.
We also investigate the effect of the conserved magnetization on
the dynamics of the spin tunnelling. We hope our studies would
bring more attention to spin tunnelling experimentally.

\section{Acknowledgments}

This work was supported by National Natural Science Foundation of
China (Grant No.: 10474008,10604009) and by Science and Technology
fund of CAEP. We thank Profs. Qian Niu and Biao Wu for stimulating
discussions.

\end{document}